\documentclass[aps, prb, twocolumn, superscriptaddress]{revtex4-1}

\usepackage{amsmath, mathtools, nccmath, amsfonts,verbatim}
\DeclareMathOperator*{\argmin}{arg\,min}
\usepackage{physics}

\usepackage{siunitx}

\usepackage{graphicx}
\usepackage{hyperref}
\hypersetup{
    colorlinks,
    linkcolor={red!50!black},
    citecolor={blue!50!black},
    urlcolor={blue!80!black}
}
\usepackage{float}
\usepackage{xcolor}

\begin{document}
\title{Neural-Network Quantum States for Periodic Systems in Continuous Space}
\date{\today}
\author{Gabriel Pescia}
\affiliation{Ecole Polytechnique Fédérale de Lausanne (EPFL), Institute of Physics, CH-1015 Lausanne, Switzerland}
\author{Jiequn Han}
\affiliation{Center for Computational Mathematics, Flatiron Institute, New York, NY 10010, USA}
\author{Alessandro Lovato}
\affiliation{Physics Division, Argonne National Laboratory, Argonne, IL 60439}
\affiliation{Computational Science Division, Argonne National Laboratory, Argonne, Illinois 60439, USA}
\affiliation{INFN-TIFPA Trento Institute of Fundamental Physics and Applications, 38123 Trento, Italy}
\author{Jianfeng Lu}
\affiliation{Departments of Mathematics, Physics, and Chemistry, Duke University, Durham, NC 27708, USA}
\author{Giuseppe Carleo}
\affiliation{Ecole Polytechnique Fédérale de Lausanne (EPFL), Institute of Physics, CH-1015 Lausanne, Switzerland}

\begin{abstract}
We introduce a family of neural quantum states for the simulation of strongly interacting systems in the presence of spatial periodicity. Our variational state is parameterized in terms of a permutationally-invariant part described by the Deep Sets neural-network architecture. The input coordinates to the Deep Sets are periodically transformed such that they are suitable to directly describe periodic bosonic systems. We show example applications to both one and two-dimensional interacting quantum gases with Gaussian interactions, as well as to $^4$He confined in a one-dimensional geometry. For the one-dimensional systems we find very precise estimations of the ground-state energies and the radial distribution functions of the particles. In two dimensions we obtain good estimations of the ground-state energies, comparable to results obtained from more conventional methods. 
\end{abstract}

\maketitle

\section{Introduction}
In recent years, the field of machine learning has seen tremendous progress in various applications of high dimensional data analysis such as image recognition, language processing, or classification tasks \cite{Berry2020a, Rao2016a}. A large portion of this progress is achieved by incorporating a-priori known structure in the data to the learning algorithm. In this way, it is possible to reduce the set of possible solutions of a learning algorithm to the relevant ones \cite{Cohen2016a, Mattheakis2019a}. Because of the omnipresence of symmetries in modern physics, the application of these techniques to computational physics has become a very active field of research that has produced promising results, especially in the area of many-body quantum physics \cite{carleo_machine_2019}. In this field, since the early research work, considerable attention was devoted to the problem of approximating ground-state wave functions of spin systems through artificial neural-networks. These representations, known as neural-network quantum states (NQS) \cite{Carleo2017a}, can encode highly-entangled wave functions \cite{deng_quantum_2017,chen_equivalence_2018,levine_quantum_2019}, and are routinely used to study correlated quantum systems with discrete degrees of freedom \cite{choo_symmetries_2018,ferrari_neural_2019,hibat-allah_recurrent_2020,choo_fermionic_2020,szabo_neural_2020,bukov_learning_2021,nomura_dirac-type_2021,astrakhantsev_broken-symmetry_2021}, often improving upon existing state-of-the-art results.

More recently, NQS have been extended to study ground-state properties of fermionic systems in continuous space, introducing deep neural network architectures that by design satisfy the Pauli exclusion principle.
The focus of these approaches has been primarily on relatively small atomic and molecular systems \cite{han2019solving, Pfau2020a, Hermann2020, Li2021a,kessler_artificial_2021,wilson_simulations_2021}, as well as nuclear physics \cite{adams_variational_2021,Gnech:2021wfn}. These applications have already shown significant improvements over more traditional, physics-inspired wave functions.  
However, for these tools to also become a compelling, more accurate alternative ab-initio approach for the prediction of electronic structure properties, and bulk properties in different phases of matter, it is crucial that they are extended to periodic systems. While progress in this direction has been already realized in the context of lattice-based NQS \cite{yoshioka_solving_2021}, an open methodological issue is the extension of continuous-space NQS to efficiently encode periodicity, while preserving other fundamental symmetries, most importantly particle- permutation invariance.

The ability to treat strongly-correlated periodic systems in continuous space is of chief importance in a wide range of condensed-matter physics, nuclear physics and quantum chemistry problems. Emblematic examples of strongly-interacting periodic systems in condensed matter are the electron gas~\cite{Dornheim2016}, the bulk of bosonic and fermionic helium~\cite{Schiff1967, Francis1970,ceperley_path-integrals_1995}, supersolids ~\cite{chester_speculations_1970,tanzi_observation_2019} and high-pressure hydrogen~\cite{holzmann_backflow_2003,Azadi2013}. In nuclear physics, the matter comprising the interior of neutron stars is typically modeled as a periodic systems of strongly-interacting protons and neutrons~\cite{Carlson2003,Piarulli:2019pfq,Lonardoni:2019ypg}. It has to be noted that neural networks have been recently employed to reduce the associated finite-size effects, extrapolating quantum Monte Carlo calculations of neutron matter and unitary gas to the thermodynamic limit~\cite{Ismail:2021nir}.

In all these problems, physics-driven wave functions combined with quantum Monte Carlo methods have played a crucial role in the understanding of key phenomena, including superfluidity, superconductivity, and crystallization. Notable examples of these wave functions include Jastrow correlators \cite{jastrow_many-body_1955} applied to a mean-field Slater determinant, which may include backflow correlations \cite{feynman_energy_1956,schmidt_structure_1981,ruggeri_nonlinear_2018}. Despite their success at describing key physical phenomena, these Ansatz state stems are not systematically improvable and typically require significant adjustments when used on systems different from the one they were originally designed for.

In this work, we introduce continuous-space NQS based on a periodic transformation of the single- and two-body coordinates of particles that are suitable for the description of periodic systems of interacting bosons. These NQS variational states are by construction permutation invariant w.r.t.\ particle exchanges. We demonstrate their flexibility in solving non-relativistic Hamiltonians, including those relevant for liquid Helium and soft Gaussian cores, for different densities and system sizes. 

\begin{figure*}[ht]
  \includegraphics[width=1.9\columnwidth]{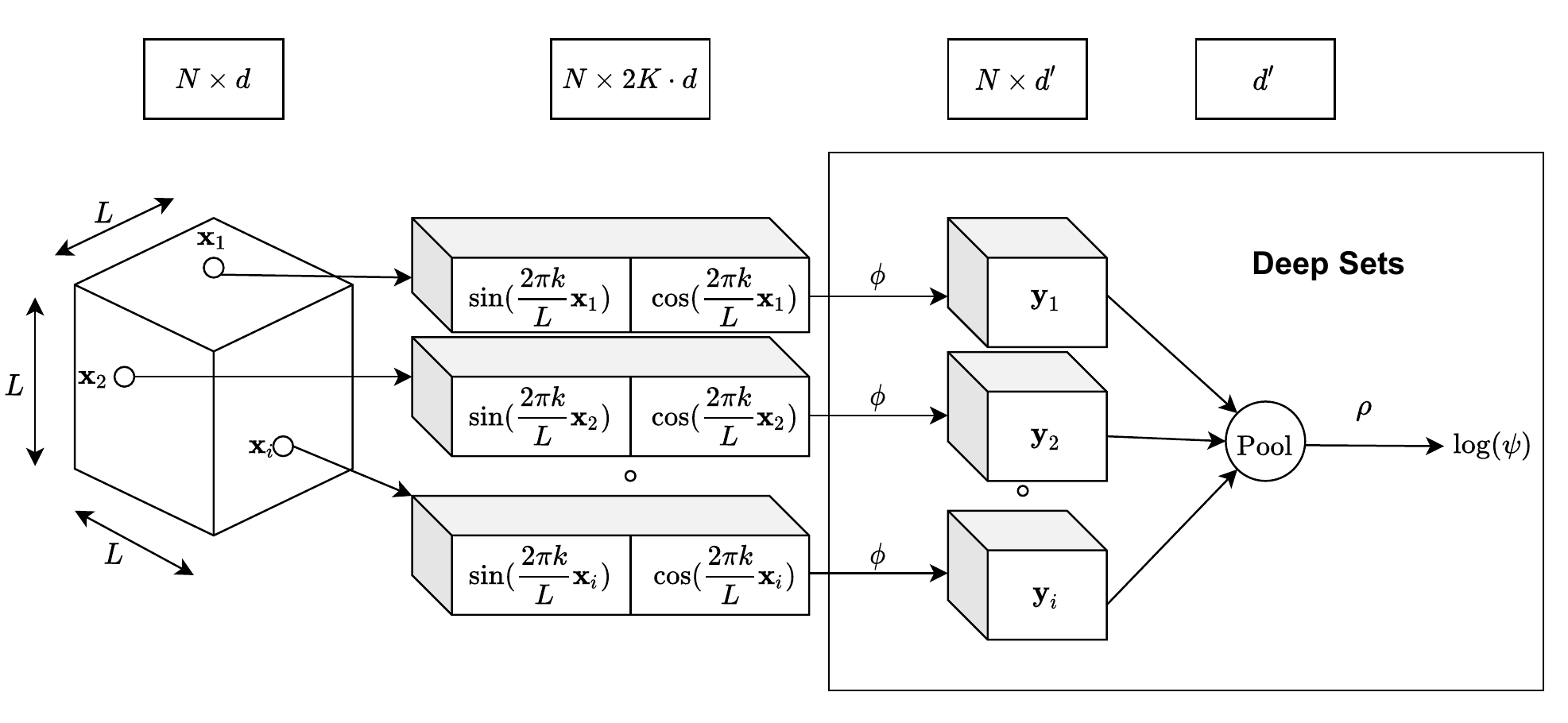}
  \caption{\label{DeepSets} Illustration of the Deep Sets neural network architecture used in this work. Starting from the left we have a simulation box in $d$ spatial dimensions of extent $L$. The single particle coordinates $\mathbf{x}_i \in \mathbb{R}^d$ are transformed to the truncated Fourier basis in Eq.\ (\ref{PBCs}) i.e.\ each single-particle coordinate is mapped to a $2K\cdot d$ dimensional vector. These periodic encodings are then fed to the DSs architecture, defined by the neural networks $\phi$ and $\rho$ and a $\mathrm{Pool}$-function. The output of the DSs architecture is the logarithm of the variational wave-function. The top row depicts the number of feature vectors and their dimensionality at each stage of the Ansatz.} 
\end{figure*}

\section{Methods}
Throughout this work we will consider the many-body Hamiltonian of $N$ interacting particles in $d$-dimensional continuous-space, given by
\begin{equation}
H = T + V = -\frac{\hbar^2}{2m} \sum_{i=1}^N \nabla_i^2 + V(x) \label{eq:hamiltonian}
\end{equation}
where $T$ and $V$ denote the kinetic and potential energy, respectively. The continuous position variable $x = (\mathbf{x}_1,...,\mathbf{x}_N)$ denotes the set of all $N$ single-particle positions $\mathbf{x}_i \in \mathbb{R}^d$. We will confine the single-particle positions to a finite box of length $L$, equipped with periodic boundaries, such that the methods introduced hereafter are well suited for the study of the bulk properties of a variety of quantum systems as well as intrinsically periodic structures such as solids.

\subsection{Periodic Boundary Conditions (PBCs)}
In the presence of PBCs, the bare single particle coordinates are not well suited as input to our variational Ansatz, since they do not reflect the periodicity of the simulation cell. The wave-function must be invariant under the translation of a single particle by a primitive lattice vector $\mathbf{L}_j = L \mathbf{e}_j$ i.e.\ $\psi_\alpha(\mathbf{x}_1,...,\mathbf{x}_i + \mathbf{L}_j,...,\mathbf{x}_N) = \psi_\alpha(\mathbf{x}_1,...,\mathbf{x}_i,...,\mathbf{x}_N)$ ($\mathbf{e}_j $ in our case denotes the Euclidean standard basis vector of $\mathbb{R}^d$) \cite{Whitlock2006a}.\\
To respect the boundary conditions we propose to map the single-particle coordinates to a truncated $L$-periodic Fourier basis \cite{han2020solving} 
\begin{eqnarray}
\mathbf{x}_i &\mapsto \left(\sin\left(\frac{2\pi k}{L} \mathbf{x}_i\right), \cos\left(\frac{2\pi k}{L} \mathbf{x}_i\right)\right)_{k=1}^K \nonumber \\
&= \mathbf{x}_i^{(K)} \in \mathbb{R}^{d\cdot 2K}. \label{PBCs}
\end{eqnarray}
Here $\sin(\mathbf{x}_i)$ and $\cos(\mathbf{x}_i)$ means, applying the trigonometric functions component-wise to the vector $\mathbf{x}_i$.
Note that $\mathbf{x}_i^{(K)}$ is invariant under the translation $\mathbf{x}_i \mapsto \mathbf{x}_i + \mathbf{L}_j$, and thus the whole wave-function Ansatz is invariant.

In order to respect the periodicity of the simulation box also in the computation of inter-particle distances, we will use $d_{\mathrm{sin}}(\mathbf{x}_i,\mathbf{x}_j) = \|\frac{L}{2}\sin(\frac{\pi}{L}\mathbf{r}_{ij})\|$ with $\mathbf{r}_{ij} = \mathbf{x}_i-\mathbf{x}_j$ as a surrogate for the ordinary Euclidean distance ($d(\mathbf{x}_i,\mathbf{x}_j) = \|\mathbf{x}_i-\mathbf{x}_j\|$) in the variational Ansatz. Distances in the potential energy will be computed using the minimum image convention \cite{Allen2004IntroductionTM}.

\subsection{Variational Ansatz}
We construct the variational wave-function Ansatz $\psi_\alpha$ as a product of a short- and a long-range part. The former is determined by requiring that the diverging behavior of the potential $V(x)$ at short inter-particle distances is compensated by the kinetic energy contribution of the wave-function, such that we obtain a finite energy contribution:
\begin{equation}
\lim_{r\rightarrow 0} \left(\frac{\nabla^2 \psi(r)}{\psi(r)} + V(x) \right) < \infty \,. \label{Cusp}
\end{equation}
This is called \textit{Kato's cusp condition} and constitutes a boundary condition for the given many-body system \cite{Kato1957b}.

The long-range part is parameterized by a neural network based on the \textit{Deep Sets} (DSs) architecture, which builds on the fact that a function $f(\mathbf{x}_1,...,\mathbf{x}_N)$ is invariant under permutations of its input iff it can be decomposed as follows \cite{Zaheer2017b,Wagstaff:2019}:
\begin{equation}
f(\mathbf{x}_1,...,\mathbf{x}_N) = \rho\left(\mathrm{Pool}\left( \phi(\mathbf{x})\right)\right)  \label{DS}
\end{equation}
where $\mathbf{x} \in \mathbb{R}^{N \times d}$ is the matrix of all single-particle coordinates $\mathbf{x}_i$. The two vector-functions $\rho(\mathbf{x})$ and $\phi(\mathbf{x})$ depend on the function $f(\mathbf{x}_1,...,\mathbf{x}_N)$ ($\phi(\mathbf{x})$ is applied row-wise to $\mathbf{x}$). The pooling operation $\mathrm{Pool}(\phi(\mathbf{x}))$ denotes the mixing of the feature vectors $\mathbf{y}_i \coloneqq \phi(\mathbf{x}_i)$ such that the output is permutational invariant with respect to the particle index (see Fig.\ \ref{DeepSets}) \cite{Zaheer2017b}. Examples of such pooling operations are the sum-pooling $\mathrm{Pool}(\mathbf{y}) = \sum_i \mathbf{y}_i$ and \textit{logsumexp}-pooling ($L\Sigma E$), defined as $L\Sigma E(\mathbf{y}) = \log\left(\sum_i \exp\left[\mathbf{y}_i\right]\right)$, which is a continuous version of the \textit{max}-pooling.

The function $\phi$ can be seen as an encoding of the single-particle coordinates into an appropriate feature space. The pooling collects the relevant features to produce a global feature vector that is then fed to the function $\rho$, which correlates the combined encodings. To have a flexible variational Ansatz applicable to a variety of different systems, the functions $\phi$ and $\rho$ are parameterized by dense feed-forward neural networks. This Ansatz can then be shown to be a universal approximator for the permutation invariant function class.

The complete neural variational Ansatz reads
\begin{eqnarray}
\psi_\alpha(\mathbf{x}_1,...,\mathbf{x}_N)&=&\prod_{i<j} \exp\left[-\frac{1}{2}u_2\left(d_{\mathrm{sin}}(\mathbf{x}_i,\mathbf{x}_j)\right)\right]\nonumber \\ &\cdot & \exp\left[\rho\left(\mathrm{Pool}\left( \phi(\mathbf{x}^{(K)})\right)\right)\right] \label{NWF} 
\end{eqnarray}
where $u_2(r)$ is chosen such that Eq.\ (\ref{Cusp}) holds and $\mathbf{x}^{(K)}$ is a matrix storing all periodized single-particle coordinates defined in Eq.\ (\ref{PBCs}).

As an alternative to the single-particle coordinates $\mathbf{x}_i^{(K)}$ it is also possible to input the periodized two-body distances between the particles $d_\mathrm{sin}(\mathbf{x_i},\mathbf{x_j}) = \|\frac{L}{2}\sin(\frac{\pi}{L}\mathbf{r}_{ij})\|$ and then pool over all indices $i,j$. This facilitates the learning procedure substantially, though at the price of higher computational complexity of the DSs.

Note that in either case (single- or two-body coordinates) the Ansatz can take an arbitrary amount of inputs. This allows us to train a small system with only a few particles, and then use the obtained optimized wave-function as an initialization for the variational Ansatz for a bigger system, containing more particles. This is only possible if the architecture for the bigger system is exactly the same as for the smaller system, thereby preventing the introduction of additional variational parameters. Consequently, if a big number of parameters is needed to describe the big system, they must already be introduced for the small system size. This has however not shown to be a great limitation in our applications.

\subsection{Optimization}
The Rayleigh-Ritz principle establishes a lower bound on the expectation value of the Hamiltonian $\bra{\psi}H\ket{\psi}/\braket{\psi}{\psi} \equiv E[\psi] \geq E_0$, allowing us to formulate a variational principle on the ground-state wave-function
\begin{equation}
\psi_0(x) = \argmin\limits_{\psi} E[\psi]\, .
\end{equation}
where $H\ket{\psi_0} = E_0\ket{\psi_0}$. 
The exact evaluation of the energy expectation value is not computationally feasible using deterministic integration methods, and we resort to sampling techniques commonly adopted in quantum Monte Carlo. The energy is estimated by accumulating samples of the local energy $E_{loc}(x) = \bra{x}H\ket{\psi_\alpha}/\bra{x}\ket{\psi_\alpha}$, where the coordinates $x$ are drawn from the probability distribution $|\psi_\alpha(x)|^2$ using the Metropolis-Hastings algorithm.

The variational principle allow us to obtain progressively better approximations to the ground-state wave function by minimizing $E[\psi_\alpha]$. The gradient components of the energy with respect to the variational parameters $\alpha_i$ are given by 
\begin{align}
    G_i=2\left( \frac{\bra{\partial_i\psi} H \ket{\psi}}{\braket{\psi}{\psi}}-E[\psi] \frac{\braket{\partial_i\psi}{\psi}}{\braket{\psi}{\psi}}\right)
\end{align}
and can be efficiently estimated through Monte Carlo sampling. The variational parameters are updated as $\delta \alpha = -\eta S^{-1}G$, where $\eta$ is the learning rate and 
\begin{align}
    S_{ij}=\frac{\bra{\partial_i\psi}\ket{\partial_j\psi}}{\braket{\psi}{\psi}}-\frac{\bra{\partial_i \psi}\ket{\psi}\bra{\psi}\ket{\partial_{j}\psi}}{\braket{\psi}{\psi}\braket{\psi}{\psi}},
\end{align}
is the Fisher-information matrix. This approach, known as the stochastic-reconfiguration (SR) algorithm~\cite{Sorella1998,Sorella2005} is equivalent to performing imaginary-time evolution in the variational manifold and it is related to the Natural Gradient descent method~\cite{Amari1998} in unsupervised learning.

\section{Results}
To test the expressiveness of the proposed variational Ansatz in Eq.\ (\ref{NWF}) and, in particular, its compatibility with the periodic transformation in Eq.\ (\ref{PBCs}) and the cusp condition in Eq.\ (\ref{Cusp}), we examine its performance first on a system of $^4$He in $d=1$ dimension interacting via a Lennard-Jones-like potential. Subsequently, we show the versatility of the Ansatz by applying in also to a cusp-less system of Gaussian cores in $d=1$ and $d=2$ spatial dimensions.

\subsection{Helium}
\begin{figure}
        \includegraphics[width=\columnwidth]{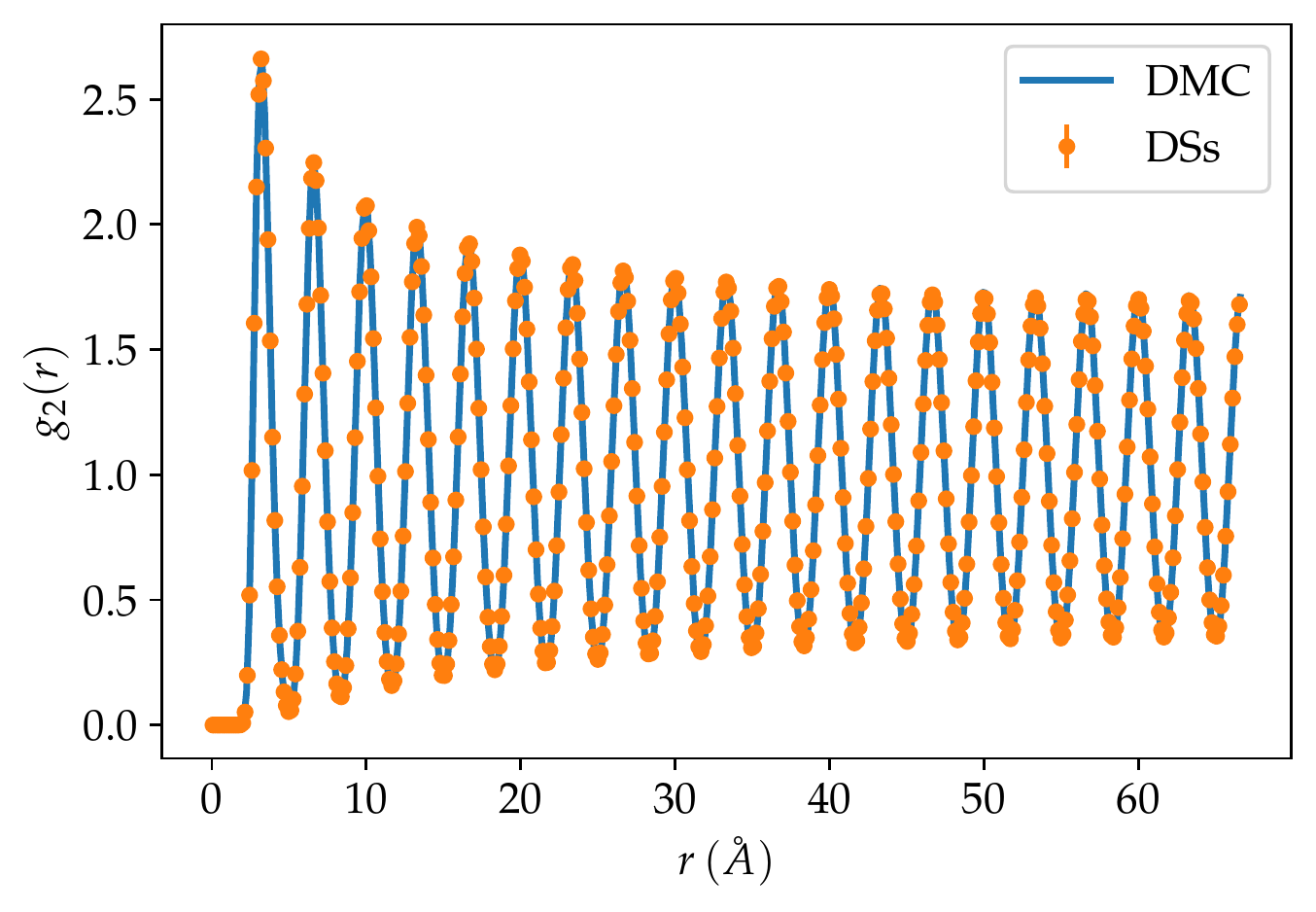} 
                \caption{DMC (solid lines) and DSs (solid points) two-body density distributions of $N=40$ particles in 1D interacting through the Aziz potential. } 
    \label{fig:g2_Helium}
\end{figure}
The inter-particle interactions of $^4$He particles in $d=1$ spatial dimension, are described by an effective two-body potential which qualitatively resembles a Lennard-Jones interaction. Here we adopt the Aziz 79 potential \cite{Bertaina2016a, Bertainaa, Aziz1977b}.

\begin{eqnarray}
V(\mathbf{x}) =\epsilon \sum_{i<j} V_{\mathrm{Aziz}}(d(\mathbf{x}_i,\mathbf{x}_j)).
\end{eqnarray} 

For large distances the potential decays rapidly while for short distances the potential resembles a Lennard-Jones potential exhibiting a $\propto r^{-12}$-divergence.
To enforce the cusp condition Eq.\ (\ref{Cusp}), this divergence needs to be compensated. It is easy to verify that the McMillan factor $u_2(r) = -\left(\frac{b}{r^5}\right)$ fulfills Eq.\ (\ref{NWF}), where we treat $b$ as a variational parameter. 

\begin{figure}
        \includegraphics[width=\columnwidth]{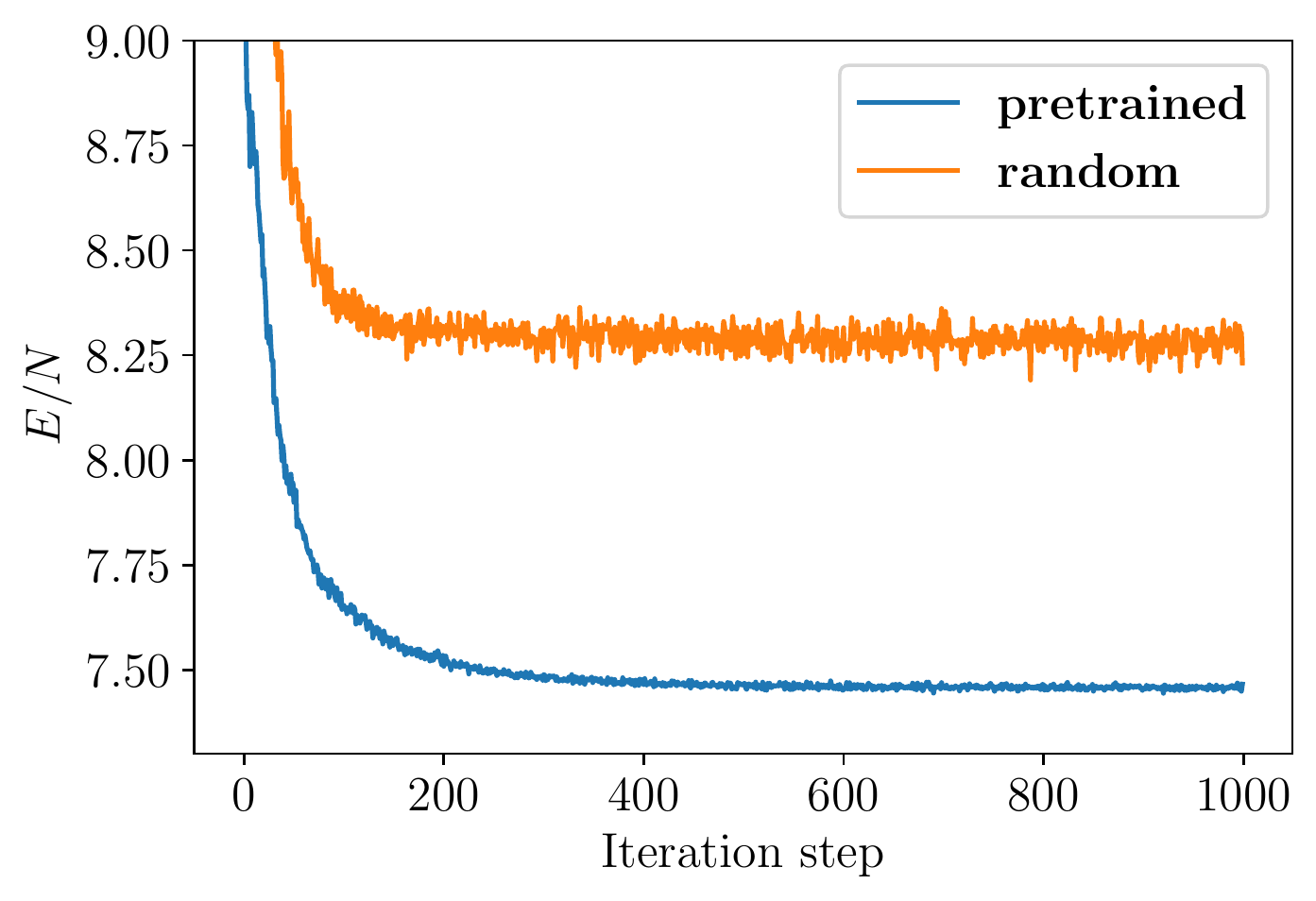}
                \caption{Convergence pattern of the DSs architecture for $N=40$ Helium particles in 1D. (blue) Shows the energy for a variational Ansatz initialized with the optimized weights for the same system at $N=10$ particles. (orange) Shows the energy for randomly initialized weights. }
    \label{fig:conv_Helium}
\end{figure}

For this system we focus on one density only ($D = 0.3 \si{\angstrom}^{-1})$ for different system sizes ($N=10,20,40$). We compare the energies obtained from the ordinary DSs model with periodized single-particle coordinates, the DSs with two-body coordinates $d_\mathrm{sin}(\mathbf{x_i},\mathbf{x_j}$), a traditional Jastrow Ansatz and diffusion Monte Carlo (DMC)~\cite{Kosztin1996} computations. Additionally we plot the two-body density distributions yielded by the optimized DSs and DMC wave-functions (Fig.\ \ref{fig:g2_Helium}).
The results are reported in Table \ref{tab:Helium}. The energies from the single-particle coordinate DSs are clearly inferior to the ones obtained by any of the other methods for $N > 10$. We suspect that the main limitation of this Ansatz is the optimization procedure rather than its representative power. To support this claim we display two training curves for $N=40$ obtained with the single-particle DSs in Fig. \ref{fig:conv_Helium}. In one case the training is started with a random initialization while in the other case, we use the optimized variational state of $N=10$ particles as initialization to the Ansatz for the $N=40$ system. Not only does the pre-trained state reach lower energies with fewer optimization steps, but the obtained energy estimate for the ground-state is also considerably lower for the pre-trained state.

In contrast to the single-particle case, the two-body DSs do not seem to have any representative or optimization problem. Note that since the variational Ansatz contains two-body coordinates in the cusp-part, the computational complexity is not increased when using two-body coordinates.
The obtained energies are close to the ones obtained from DMC and consistently lower than the Jastrow energies (or within statistical error). Also the two-body density distribution fits the DMC results almost perfectly, in particular also at large inter-particle distance (see Fig.\ \ref{fig:g2_Helium}). Note the strong oscillations in the two-body distribution, indicating that two particles can not come closer to each other than the Aziz core. Since in one-spatial dimension the particles cannot go around each other, a crystalline-like structure emerges.  

\begin{table}
\begin{center}
\begin{tabular}{ c  c  c  c  c } 
 \hline\hline
 $N$ & DeepSets & \parbox{2cm}{DeepSets \\ two-body} & Jastrow & DMC \\ [0.5ex] 
 \hline
 10 & 7.272(7) & 7.269(5) & 7.273(9) & 7.269(3) \\ 
 \hline
 20 & 7.391(9) & 7.387(3) & 7.424(9) & 7.375(2) \\
 \hline
 40 &  7.456(3) & 7.403(4) & 7.406(5) & 7.404(3)\\
 \hline\hline
\end{tabular}
\caption{Energy per particle $E/N$ for $d=1$ dimensional $^4$He. Displayed are three different system sizes ($N=10,20,40$) at $D = 0.3\si{\angstrom}^{-1}$ obtained variationally by a DSs architecture with single- and two-body coordinates, a Jastrow Ansatz and by means of DMC.}
\label{tab:Helium}
\end{center}
\end{table}

\subsection{Gaussian Cores}
\begin{figure*}
        \includegraphics[width=\columnwidth]{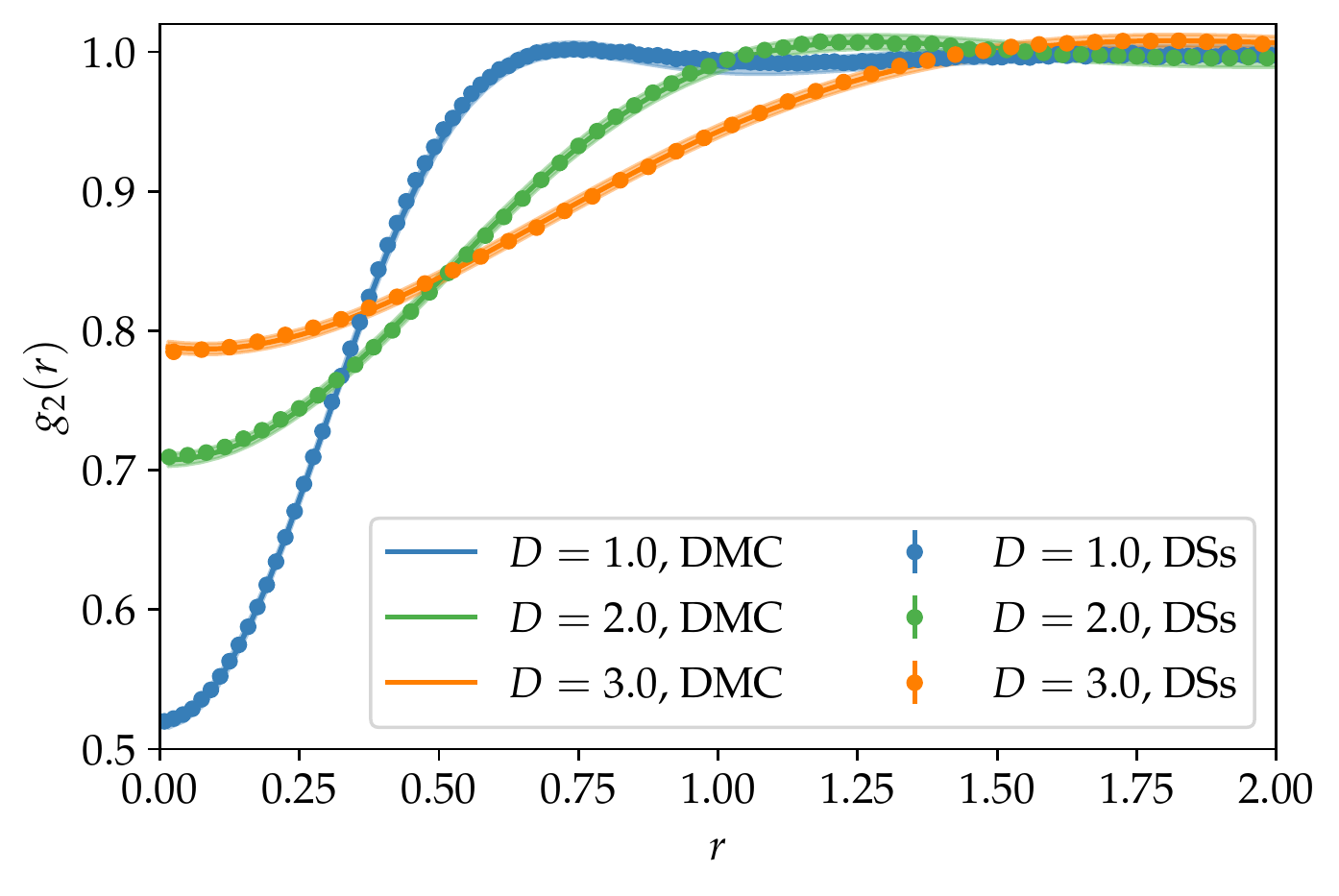} 
        \includegraphics[width=0.985\columnwidth]{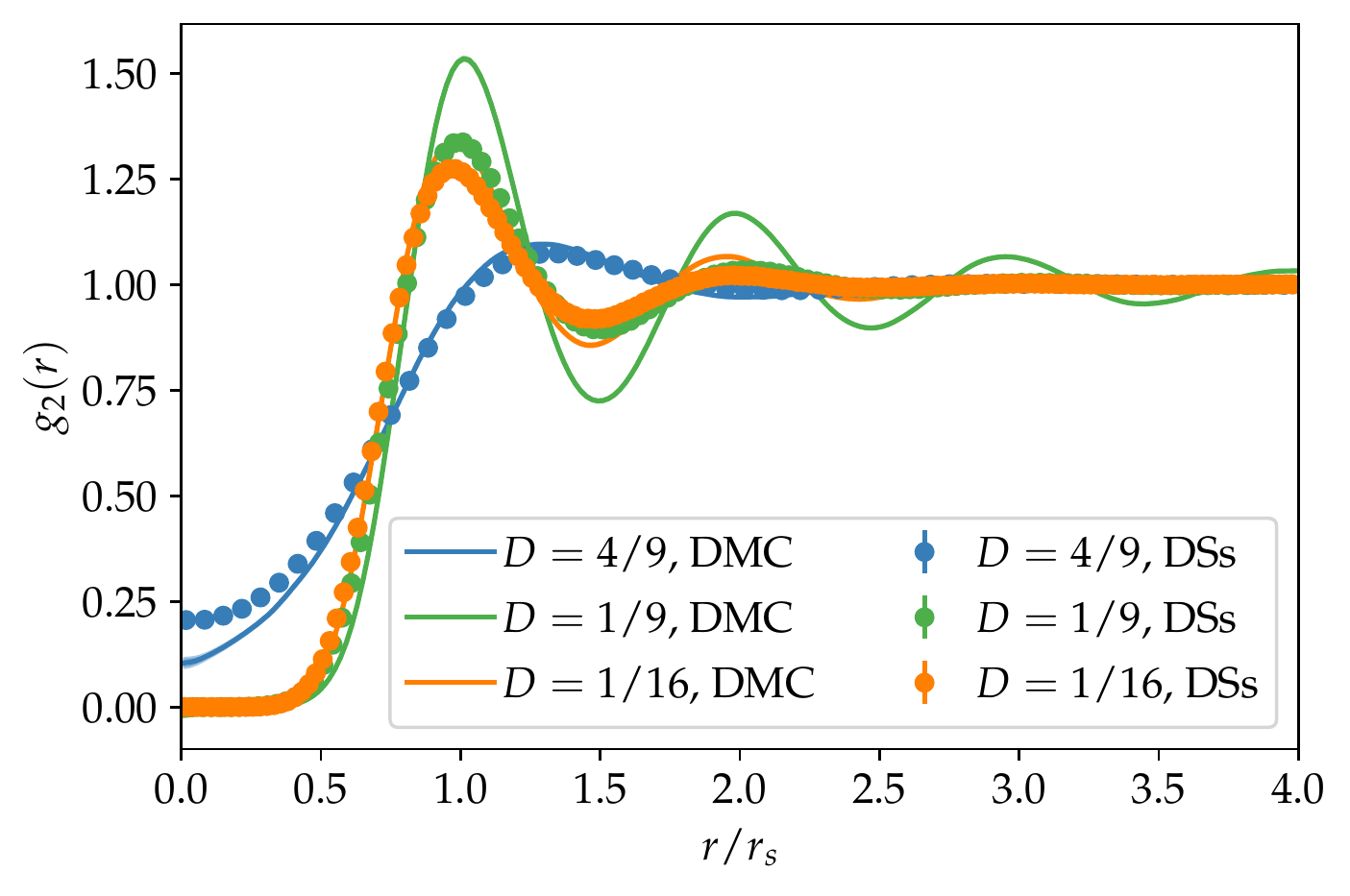} 
                \caption{(Left) DMC (solid lines) and DSs (solid points) two-body density distributions of $N=50$ particles in 1D at different densities interacting through the Gaussian potential. (Right) The two-body density distribution of $N=64$ particles in 2D at different densities interacting through the Gaussian potential. } 
    \label{fig:g2_Gaussian}
\end{figure*}

The Hamiltonian of the Gaussian cores is given by Eq.\ (\ref{eq:hamiltonian}) with a Gaussian interaction potential \cite{Kroiss2016a}
\begin{equation}
V(\mathbf{x}) = \epsilon \sum_{i<j} \exp\left[-\frac{d(\mathbf{x}_i,\mathbf{x}_j)^2}{2\sigma^2} \right].
\end{equation}
Since the interaction does not exhibit divergent behaviour in the zero-distance limit, we take $u_2 = 0$ in Eq.\ (\ref{NWF}) and the variational Ansatz solely consists of the DSs architecture.

We again benchmark the energies obtained with this NQS with variational energies given by DMC computations and a traditional Jastrow Ansatz. Additionally we plot the two-body density distributions yielded by the optimized DSs and DMC wave-functions (Fig.\ \ref{fig:g2_Gaussian}).

For the one-dimensional system we studied two different system sizes ($N=20,50$) at three different densities ($D = \frac{N}{L} = 1,2,3$) with $\epsilon=2, \sigma = 2^{-1/2}$. As can be seen in Table \ref{tab:1DGaussian} the two variational energies obtained from DSs and Jastrow as well as the DMC energies lie within statistical error of each other. We also obtain almost perfect agreement for the radial distribution functions from the DSs and DMC, indicating that our variational state, appropriately describes the ground-state of the system. An advantage of the DSs architecture over the Jastrow Ansatz is, that the computational complexity only scales as $\mathcal{O}(N)$ for one evaluation of the Ansatz, rather than $\mathcal{O}(N^2)$ as is the case for the Jastrow.

In the case of two spatial dimensions, we compare the same three models (DSs, Jastrow, DMC) for two different system sizes ($N=32, 64$) at three different densities ($D = \frac{N}{L^2} = 4/9,1/9,1/16$) with $\hbar^2/(m\epsilon \sigma^2) = 1/30$. Instead of single-particle coordinates, we use the periodized two-body distance between the particles as input to the DSs. We find that the DSs and Jastrow are again very close. However the DMC computations yield smaller energies. The energy difference is notable in the radial distribution function. In particular for the density $D=1/9$, the DSs cannot reproduce the oscillating behaviour visible in the DMC result. This might indicate a limitation of the expressiveness of our variational Ansatz.\\

\begin{table}
\begin{center}
\begin{tabular}{ c  c  c  c  c  c } 
 \hline\hline
 N & $D$ & DeepSets & Jastrow & DMC \\ [0.5ex] 
 \hline
 20 & 1.0 & 1.2373(4) & 1.2373(1) & 1.2371(1)\\ 
 \hline
 20 & 2.0 & 2.9223(5) & 2.9224(1) & 2.9222(1) \\
 \hline
 20 & 3.0 & 4.6445(7) & 4.6447(1) & 4.6446(2) \\
 \hline
 50 & 1.0 & 1.2402(5) & 1.2408(1) & 1.2390(4) \\ 
 \hline
 50 & 2.0 & 2.9286(3) & 2.9310(1) & 2.9280(2) \\
 \hline
 50 & 3.0 & 4.6557(1) & 4.6597(1) & 4.6553(2)\\
 \hline\hline
\end{tabular}
\caption{Energy per particle $E/N$ for $d=1$ dimensional Gaussian cores. Displayed are two different system sizes ($N=20,50$) and three different densities ($D=1,2,3$) obtained variationally by a DSs architecture and a Jastrow Ansatz and by means of DMC.}
\label{tab:1DGaussian}
\end{center}
\end{table}

\begin{table}
\begin{center}
\begin{tabular}{ c  c  c  c  c  c } 
 \hline\hline
 N & $D$ & \parbox{2cm}{DeepSets \\ two-body}  & Jastrow & DMC \\ [0.5ex] 
 \hline
 32& 4/9 & 1.01707(4) & 1.02020(3) & 1.01355(4) \\ 
 \hline
 32 & 1/9 & 0.08864(5) & 0.08895(2) & 0.08546(4) \\
 \hline
 32 & 1/16 & 0.02545(5) & 0.02549(1) & 0.02434(1)\\
 \hline
 64 & 4/9 & 1.01621(3) & 1.02661(4) & 1.01429(6) \\ 
 \hline
 64 & 1/9 & 0.08905(3) & 0.08994(2) & 0.08519(4) \\
 \hline
 64 & 1/16 & 0.02573(5) & 0.02580(2) & 0.02443(5) \\
 \hline\hline
\end{tabular}
\caption{Energy per particle $E/N$ for $d=2$ dimensional Gaussian cores. Displayed are two different system sizes ($N=32,64$) and three different densities ($D=4/9,1/9,1/16$) obtained variationally by a DSs architecture and a Jastrow Ansatz and by means of DMC.}
\label{tab:2DGaussian}
\end{center}
\end{table}

\subsection{Influence of Network Architecture}

For all the computations we used an MLP with one and two hidden layers, respectively, as parameterization for the functions $\phi$ and $\rho$ of Eq.\ (\ref{DeepSets}). We chose the $L\Sigma E$-pooling and sum-pooling, for single- and two-body input, respectively. The non-linearity in the neurons of the MLPs was chosen to be the $\mathrm{gelu}$-activation~\cite{Hendrycks2016} as a differentiable surrogate for the $\mathrm{ReLu}$-activation.\\

\begin{figure}
        \includegraphics[width=\columnwidth]{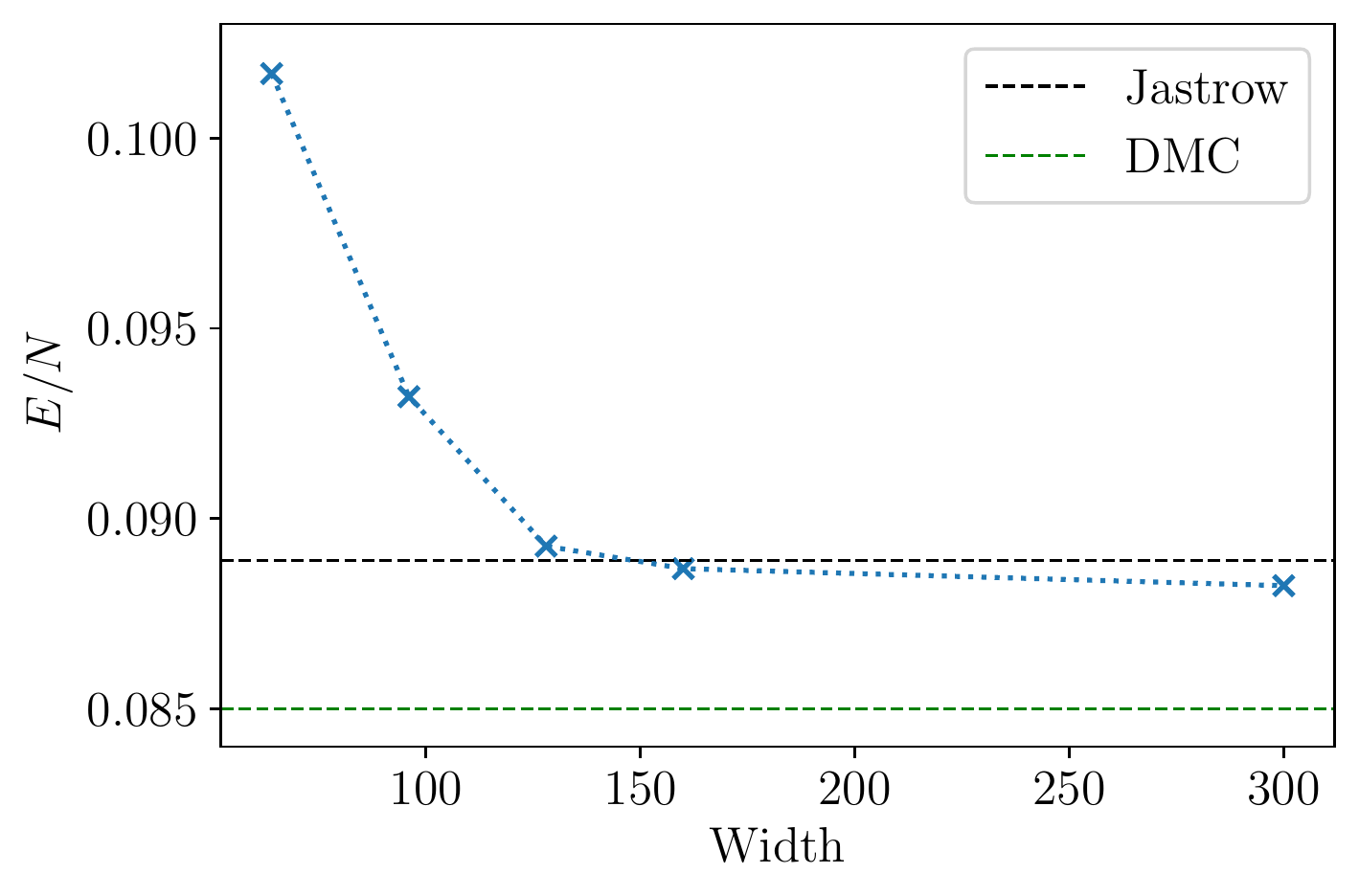}
                \caption{Final energy of training the DSs with single-particle input for the two dimensional Gaussian core model with $N=32$ particles at a density $D=1/9$ as a function of the width of the neural networks used. The horizontal lines correspond to the Jastrow and DMC results, respectively.}
    \label{fig:conv_gaussian2d}
\end{figure}

Increasing the number of hidden layers in either of the two MPLs has not increased the accuracy of the results. However increasing the width of the layers has a significant impact on the quality of the results, in particular for the $d=2$ dimensional Gaussian cores. In Fig. \ref{fig:conv_gaussian2d} we show the training curves and final energies for $N=32$ particles with density $D=1/9$, interacting via a Gaussian potential in two spatial dimensions. As an Ansatz the single-particle DSs are used with $K=10$ (see Eq.\ \ref{PBCs}). While for $<128$ neurons per layer, the ordinary Jastrow Ansatz has an edge over the DSs, our NQS starts to outperform the Jastrow Ansatz for wider layers. The DMC energies do not seem to be reachable by simply increasing the size of the layers, indicating the necessity of refinements to the current architecture.

\section{Discussion}
We have introduced an extension of NQS to systems in continuous space and with intrinsic spatial periodicity. The approach we discuss in this work is specialized to bosonic degrees of freedom, and it is based on permutationally equivariant states based on Deep Sets, featuring periodicity under spatial translations. 
We have successfully applied our periodic DSs neural quantum state to one dimensional Gaussian cores, confined to a periodic simulation cell, and showed that the obtained system properties typically improve upon traditionally used wave functions based on two-body Jastrow correlators, and are in excellent agreement with exact Diffusion Monte Carlo results. An advantage of the NQS-based approach is that it can be applied to different systems without significant modifications. Specifically, using the exact same Ansatz (up to modifying the input) to model interacting $^4$He in one spatial dimension, we observe excellent agreement with existing state-of-the-art methods. Modeling the Helium system is considerably more difficult than the Gaussian cores because of the rapidly diverging potential at short distances. For the case of two spatial dimensions, we obtain energies that are marginally better than the ordinary Jastrow Ansatz. We show that the energy estimate can be improved by increasing the width of the neural networks, however we are not able to reach the DMC energies even with the largest network used.

An important result for the optimization of our variational state is that the optimized Ansatz for small systems can very well be used as an initialization for larger systems (known as transfer learning). However it remains to be seen, if for even larger systems than were presented in this article, a higher number of parameters is necessary to capture all the relevant correlations between the particles. If so, these parameters must already be present in the training of the small system when applying the transfer learning method above, potentially making the training of small systems less efficient.

Concluding, we showed that our NQS respects the periodicity of the simulation cell as well as the permutation invariance of the wave-function w.r.t.\ particle exchange, while still yielding high-precision results. It can be envisioned that the current architecture is altered to better account for the locality of the interaction potential as well as further relevant symmetries of the problem.  We suspect that a more flexible Ansatz than DSs will help systematically improve the accuracy in the two dimensional case, and we leave this aspect for future studies. Also the architecture might be used to model fermionic systems, which ordinarily are modelled as product of a symmetric and anti-symmetric part. The DSs can provide the symmetric component of such an Ansatz.

\section{Acknowledgement}
The authors acknowledge several insightful discussions with David Ceperley, Markus Holzmann, and Saverio Moroni. The numerical tools used in this are based on the open-source software NetKet\cite{carleo_netket:_2019, Vicentini2021} version 3.0. The present research is supported by the Swiss National Science Foundation under Grant No. 200021\_200336, by the U.S. Department of Energy, Office of Science, Office of Nuclear Physics, under contract DE-AC02-06CH11357, the NUCLEI SciDAC program (A.L.), the DOE Early Career Research Program, and Argonne LDRD awards. Part of the calculations presented in this work were performed through a CINECA-INFN agreement that provides access to resources on MARCONI at CINECA. This collaboration originates from a SQuaREs program hosted by American Institute of Mathematics, we thank AIM for their hospitality.

\bibliography{DeepSets.bib}

\end{document}